\documentclass{article}
\usepackage{graphicx}
\begin{document}
\title{Rigid body motion in special relativity}
\author{Jerrold Franklin\footnote{Internet address:
Jerry.F@TEMPLE.EDU}\\
Department of Physics\\
Temple University, Philadelphia, PA 19122-6082}
\date{June 7, 2012}
\maketitle

\begin{abstract}
We study the acceleration and collisions of rigid bodies in special relativity.
After a brief historical review,  we give a physical definition of the term `rigid body' in relativistic straight line motion.
We show that the definition of  `rigid body' in relativity differs from the usual classical definition, so there is no difficulty in dealing with rigid bodies in relativistic motion.
We then describe
\begin{enumerate}
\item The motion of a rigid body undergoing constant acceleration to a given velocity.
\item  The acceleration of a rigid body due to an applied impulse.
\item Collisions between rigid bodies. 
\end{enumerate}
\end{abstract}
\maketitle

\section{Introduction}

How can we write about rigid bodies in special relativity when some authorities deny their existence in special relativity?
For instance, Pauli\cite{pauli} wrote ``the concept of a {\it rigid body} has no place in relativistic mechanics," while 
Panofsky and Phillips\cite{pp} state that special relativity ``precludes the existence of the `ideal rigid body' ."
Most other textbooks do not mention the words `rigid body' in connection with special relativity.
Yet, in his 1905 paper\cite{ae}, Einstein writes the phrases ``Let there be given a stationary rigid rod ...", and
``We envisage a rigid sphere...", and four years later Born\cite{born} postulated conditions for rigid body motion in relativity.
  Thus rigid bodies are at the heart of special relativity, yet some authorities deny their existence.

Can we resolve these statements?  Although the previous quote of Pauli is often referred to, he went on to add
``it is nevertheless useful and natural to introduce the concept of a {\it rigid motion} of a body."  What does he mean by this? 
Pauli's (and others\cite{{gh},{fn}}) objection to use of the term `rigid body' in special relativity was that the general motion, including rotation, of a rigid body could not be described in relativity.  That seems to be asking too much of it, since special relativity includes only Lorentz transformations and no (nonstatic) rotational transformations for anything.
We shall thus discuss the motion of a rigid body only in translational motion, and Pauli did countenance that.

The objection of Panofsky and Phillips to a rigid body
is that ``its ends would move simultaneously as observed from any frame".
However, we will show below that the ends of a relativistic rigid rod do not move simultaneously as observed from any frame, removing the objection of Panofsky and Phillips.

The problem is also raised that if a truly rigid body were kicked at one end, the other end would move instantly rather than at a retarded time.
This is not only a relativistic objection.  The fact that an electromagnetic signal could not propagate faster than $c$ was shown long before the advent of relativity.  Since a rod is held together by electromagnetic forces, the simultaneous motion of the right end if the left end were kicked is ruled out on classical grounds, because of the need to use the retarded time.  Actually, of course, $c$ is an unrealistically fast upper limit to the speed of motion in a material rod. The actual transmission speed of an impulse is really governed by the speed of sound in the rod, which is orders of magnitude less than $c$.  Even so, the physical abstraction of a rigid rod with seemingly infinite speed of transmission of impulse is a useful and much used concept in classical physics.  

One more consideration in dealing with rigid bodies is that ideal rigid bodies actually do exist in nature.  In the Mossbauer effect,  the entire absorbing crystal moves impulsively at the same instant. 
This is a cooperative quantum effect, and there is nothing in relativity forbidding it.

Perhaps because of the negative comments on rigid bodies in special relativity, there have been relatively few publications[7-11] 
on translational relativistic rigid body motion in the years since the original Born paper.  We\cite{jf} have also discussed the constant acceleration of a rigid body in connection with the motion of Bell's spaceships\cite{bell}. 
In this paper, we extend those treatments, and give specific trajectories for constant acceleration, impulsive acceleration, and collisions between rigid bodies.

\section{Definition of a `relativistic rigid body'}

In classical (prerelativistic) dynamics, the motion of a rigid body is generally defined as preserving the dimensions of the body during any motion of the body.
There are two problems with this definition.   First, any actual physical object will have elastic properties, so there must be some distortion during accelerated motion.  Second, due to the finite velocity of sound in any real object, one end of a rigid rod will not move until a short time after the other end is struck.  
These difficulties are generally dispensed with by
assuming that the body is so rigid that the elastic deformation can be ignored, and the speed of sound so fast that the initial delay in the motion of the other end can also be neglected.  This leads to the abstraction of an `ideal rigid body' that is used in all the books and papers treating classical rigid body motion.

These two approximations are also necessary in the relativistic treatment of rigid bodies.  The additional approximation of neglecting the time delay due to the finite velocity of light is no problem since the relativistic time delay is so much shorter than the delay due to the speed of sound, already neglected in classical dynamics.
An additional objection is often made that the `relativistic length' of a moving object changes as its velocity increases.  This would violate the classical definition that rigid body motion preserves the dimensions of a body during any motion of the body.
This is an example of how using a prerelativistic definition for a relativistic phenomenon leads to confusion.  

In fact the proper relativistic definition of a rigid body turns the classical definition on its head.  If an object retained its length while moving, its length would increase in its rest system.  Consequently, we take as our definition of a rigid body that {\em a rigid body retains its rest frame dimensions  while in translational motion.}
This requires a moving rigid body to change its `relativistic length' in any frame in which it is moving.

\section{Constant acceleration}

This section is based on the derivation for constant acceleration in Ref.\cite{jf}.
We consider the motion of a rigid rod of length $L_0$ that starts from rest in a Lorentz system S. 
We assume an acceleration that is constant in time so that we can find explicit trajectory equations for the motion.
Thus each point on the rod undergoes a constant acceleration in its instantaneous rest system S$'$.
By `instantaneous rest system', we mean a Lorentz system moving at constant velocity in which that point on the rod is momentarily at rest.
 We show below that, in order to keep a constant length in its rest system, the front and back ends of the rod must have different constant accelerations, 
$\bf a'_F$ and $\bf a'_B$, in the rest system.  

As the rod's velocity increases in the frame S, an acceleration  $a'$ of any point on the rod in its rest system is related to the acceleration $a$ in frame S where that point is moving with velocity $v$ by (using units with $c=1$)
\begin{equation}
a'=\gamma^3 a=(1-v^2)^{-\frac{3}{2}}\frac{dv}{dt}.
\end{equation}
This equation follows from Eq.\ (14.26) of Ref.\cite{jftext} for a point at rest in system S$'$.
Since $a'$ is constant, we can integrate this equation to get
\begin{equation}
a't=\int(1-v^2)^{-\frac{3}{2}}dv=\frac{v}{\sqrt{1-v^2}}.
\label{eq:ap}
\end{equation}
We solve this for v, getting
\begin{equation}
v=\frac{a't}{\sqrt{1+a'^2t^2}}=\frac{dx}{dt}.
\label{eq:vel}
\end{equation}
One more integration gives
\begin{equation}
x=x_0+\int_0^t\frac{a'\bar{t}d\bar{t}}{\sqrt{1+a'^2\bar{t}^2}}\nonumber\\
=x_0+\left(\sqrt{1+a'^2t^2}-1\right)/a'.
\label{eq:deriv}
\end{equation}  

The equation of motion of each end of the rod is given by Eq.\ (\ref{eq:deriv}) as
\begin{eqnarray}
x_F&=&L_0+\left(\sqrt{1+a'^2_F t_F^2}-1\right)/a'_F
\label{eq:xf}\\
x_B&=&\left(\sqrt{1+a'^2_Bt_B^2}-1\right)/a'_B,
\label{eq:xb}
\end {eqnarray}
where $t_F$ and $t_B$ are the times at which the front ($x_F$) and back ($x_B$)  ends of the rod are measured.

Rigid body motion for the rod means keeping the distance between the ends of the rod constant at $L_0$ in their mutual rest system.  In order to transform to the rest system of the rod, we have to know $x_F$, $x_B$,
$t_F$, and $t_B$ when each end has the same velocity in S.  We can do this by using the relations
\begin{equation}
t=\gamma v/a'\quad{\rm and}\quad\gamma=\sqrt{1+a'^2t^2},
\end{equation}
which follow from Eqs.\ (\ref{eq:ap}) and (\ref{eq:vel}) above. 
 Then, we have
 \begin{eqnarray}
x_F&=&L_0+(\gamma-1)/a'_F\nonumber\\
x_B&=&(\gamma-1)/a'_B
\label{eq:xlt}
\end {eqnarray}    
for the location of each end of the rod 
when they have the same velocity $v$.  The two times $t_F$ and $t_B$ are now different.  These times are given by
 \begin{eqnarray}
t_F&=&\gamma v/a'_F\nonumber\\
t_B&=&\gamma v/a'_B.
\label{eq:tlt}
\end{eqnarray} 

The condition that the distance between the ends in the rest system be fixed at $L_0$ can be imposed by Lorentz transforming their difference 
$\Delta x=x_F-x_B$
 in system S to the rest system.  
The space and time differences for the two ends follow from Eqs. (\ref{eq:xlt}) and (\ref{eq:tlt}):
 \begin{eqnarray}
\Delta x&=&L_0+(\gamma-1)\delta
\label{eq:dx}\\
\Delta t&=&\gamma v\delta,
\end{eqnarray} 
where 
\begin{equation}
\delta=\frac{1}{a'_F}-\frac{1}{a'_B}.
\end{equation}
The Lorentz transformation to the rest frame is
\begin{eqnarray}
L_0&=&\Delta x'=\gamma(\Delta x-v\Delta t)\nonumber\\
&=&\gamma[L_0+(\gamma-1)\delta-v^2\gamma\delta]\nonumber\\
&=&\gamma L_0+(1-\gamma)\delta.
\end{eqnarray}
This equation has the solution
\begin{equation}
L_0=\delta=\frac{1}{a'_F}-\frac{1}{a'_B},
\label{eq:lfb}
\end{equation}
so the acceleration of the back end of the rod is related to that of the front end by
\begin{equation}
a'_B=\frac{a'_F}{1-a'_F L_0}.
\label{eq:gl}
\end{equation}

Thus there is a fixed relation between the constant accelerations of the two ends of the rod in its instantaneous rest system.  Maintaining these different rest frame accelerations for each end will keep the rest frame distance, $L_0$, between them constant. 
The variation in acceleration also holds for any point on the rod, with
its rest frame acceleration given  by $a'_B$ in 
Eq.\ (\ref{eq:gl}) with $a'_F$ being the acceleration of the front end and $L_0$ representing the $x$ distance from the front end. 
 
We see that in order to keep a body rigid in its rest frame, the acceleration has to vary throughout the body in a specific way.
Although the acceleration varies, there will be no strain because this varying acceleration preserves the rest frame dimensions of the body.  Any stress in the body will not be appreciably different than the stress induced by non-relativistic acceleration of a rigid body.  Also, it does not matter where on the rigid body the impetus for acceleration acts.  The accelerated motion is a cooperative process with the acceleration of any part of the rigid body being specified by Eq.\ (\ref{eq:gl})

Although the two times $t_B$ and $t_F$ are different in the frame S where the rod is moving, the rest frame times $t'_B$ and $t'_F$ at which the ends of the rod are measured are equal.  This is shown by
the Lorentz transformation
\begin{eqnarray}
t'_F-t'_B&=&\Delta t'=\gamma(\Delta t-v\Delta x)\nonumber\\
&=&\gamma[\gamma v\delta-vL_0-v(\gamma-1)\delta]\nonumber\\
&=&\gamma v\delta-\gamma vL_0=0.
\label{eq:tpz}
\end{eqnarray}

The results above give the motion of the ends of a rigid rod undergoing continuous constant acceleration.  We now relate this to a rod that undergoes constant acceleration
from rest that ends when the rod reaches a final velocity $V$.  
We see from Eq.\ (\ref{eq:tpz}) that the acceleration stops at the same time for each end in the rest frame.
However,   Eq.\ (\ref{eq:tlt}) shows that, in frame S, the back end will 
reach the velocity $V$ at a time $T_B=\gamma V/a'_B$, which is earlier than the time $T_F=\gamma V/a'_F$ at which the front end reaches velocity $V$. 
This means that,  starting at $T_B$, the back end will move at constant velocity, while the 
front end continues to accelerate until $T_F$, at which time each end will continue with the same velocity $V$.

The motion of the rod in frame S is shown as the solid trajectory in Fig.\ 1. 
\\

\begin{center}
\includegraphics[width=4in]{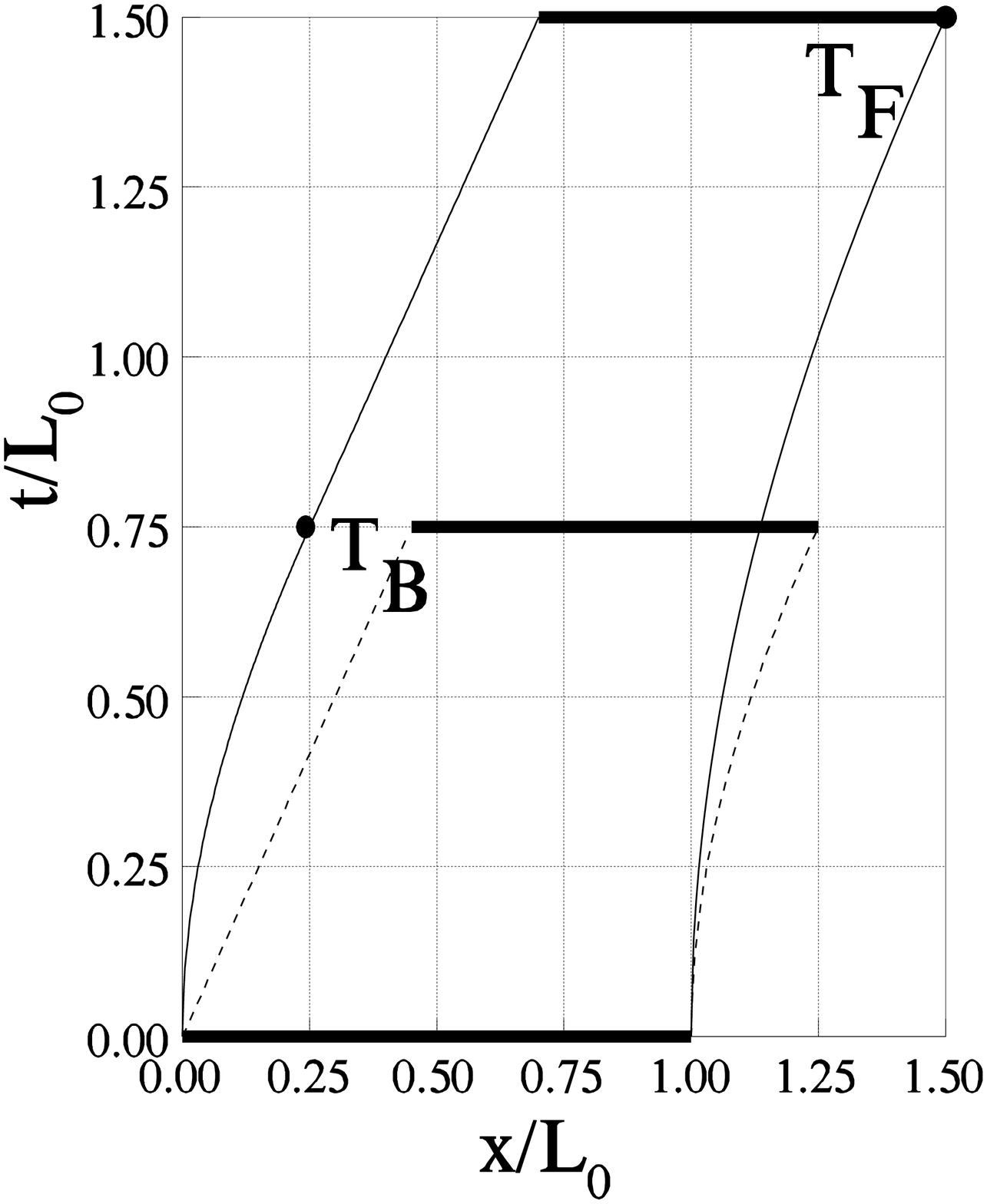}
\end{center}
\noindent
{\bf Fig.1:}  Constant acceleration of a rigid body.  The solid curve is the trajectory for continuous acceleration.  The dashed curve is for impulsive acceleration.  The time ${\bf T_B}$ on the solid curve represents the end of acceleration for the back end of the rod, and ${\bf T_F}$ for the front end.\\   
\\
The figure represents the space-time curve for acceleration in frame S from rest to a final velocity $V=0.6$, for which $\gamma=1.25$.
We have chosen the rest frame accelerations to be $a'_F=1/(2L_0)$ and $a'_B=1/L_0$, which are consistent with Eq. (\ref{eq:lfb}).
The acceleration continues until each end of the rod reaches velocity $V$, which occurs at equal times in the rest system, but at the unequal times
$T_F$ and $T_B$ shown on the figure.  At the time the acceleration stops in the rest system,  the front end of the rod is at a position $X_F$ and the back end is at $X_B$.  
The difference $X_F-X_B$ is given by Eq.\ (\ref{eq:dx}) 
to be $\gamma L_0=(5/3) L_0$.
This length is at different times in system S, but it would be the measured length if observers in S made the length measurement when told to by passengers at the front and back ends of the rod. 

After the acceleration stops in the rest frame,
the back end of the rod travels in frame S at constant velocity $V$ from time $T_B$ to $T_F$, while the front end will continue to accelerate until time $T_F$.  The  length (When we use the word `length' or the symbol $L$ without a qualifier, we mean the difference $x_F-x_B$ measured at equal times in system S.) of the rod
decreases  to 
\begin{equation}
L=\gamma L_0 -V(T_F-T_B)=\gamma L_0-\gamma V^2 L_0=L_0/\gamma, 
\end{equation}
where we have used Eq. (\ref{eq:tlt}) for the time difference ($T_F-T_B$).
At time $T_F$,
both ends of the rod will have the same velocity $V,$ and they will continue to move at that constant velocity.  At any time after $T_F$, the rod's length, measured at equal times in frame S, remains a constant 
length $L=L_0/\gamma$, the usual `Lorentz contraction'.

During the accelerated motion, the distance between the ends of the rod measured at equal times 
is given until time $t_B$ by the difference 
\begin{equation}
L=x_F-x_B=\sqrt{t^2+1/a'^2_F }-\sqrt{t^2+1/a'^2_B },\quad 0\le t\le t_B,
\label{eq:la}
\end{equation}
where we have used Eqs.\ (\ref{eq:xf}) and (\ref{eq:xb}) for $x_F$ and $x_B$.
From time $t_B$ until time $t_F$, the distance between the ends  is given by the difference of $x_F$ as given by 
Eq. (\ref{eq:xf}) and $x_B$ given by $x_B=V(t-T_B)+X_B$.  After some algebra, this results in
\begin{equation}
L=\sqrt{t^2+1/a'^2_F}-Vt.\quad t_B\le t\le t_F,
\label{eq:lb}
\end{equation}

Although the motion described above keeps the rest frame length of the rod constant, we see that the distance between the ends of the rod, measured in system S at the same time for each end, will decrease.  This decrease is shown in Fig.\ 2, which is a plot of Eqs. (\ref{eq:la}) and ({\ref{eq:lb}).  The equal time length continually decreases from $L_0$ to $L_0/\gamma$ when each end has the final constant velocity $V$.  We see that while the classical definition of a rigid body requires it to have a constant length while accelerating, the relativistic definition requires its length to change.
\begin{center}
\includegraphics[width=4in]{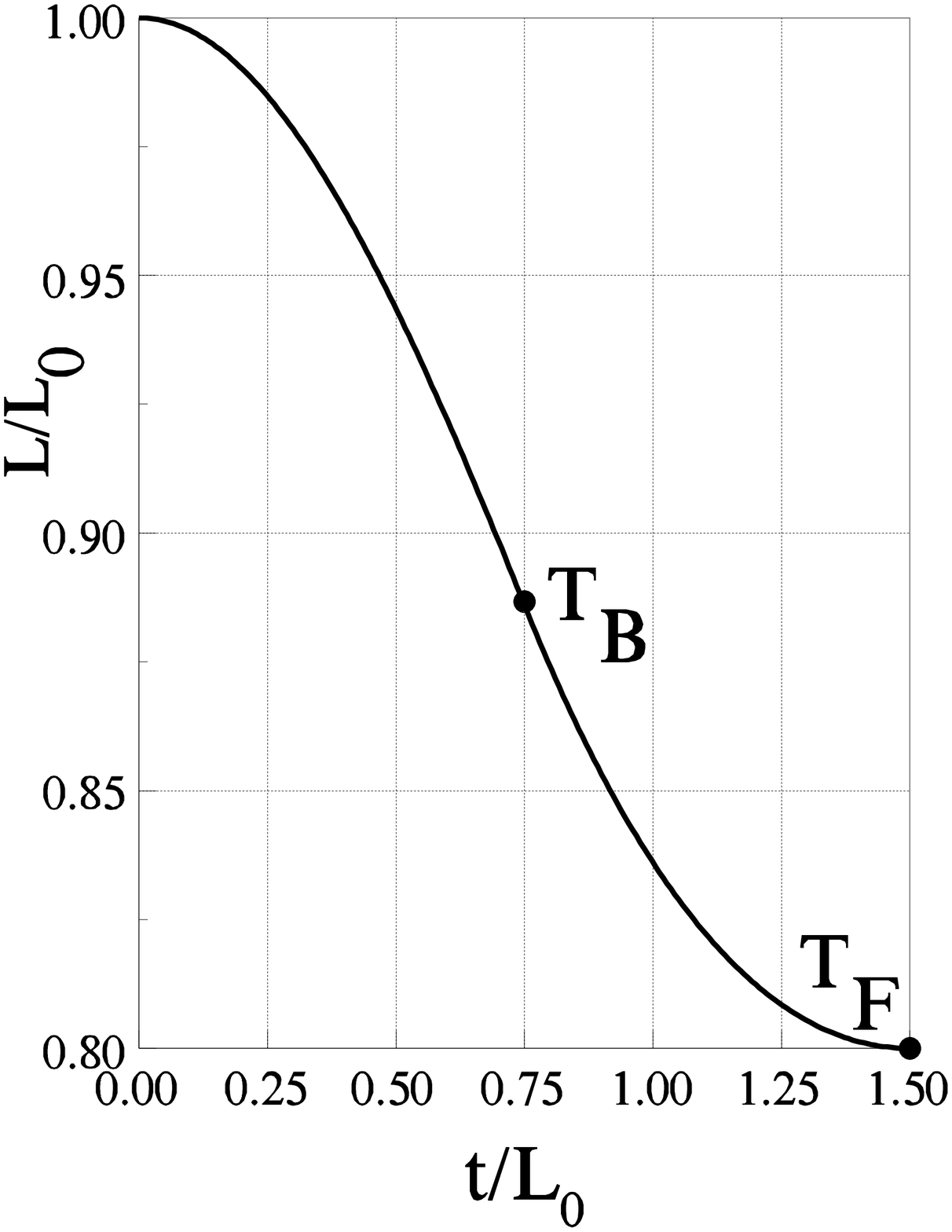}
\end{center}
\noindent
{\bf Fig. 2:}  Equal time length of an accelerating rigid rod.  
The length decreases from $L_0$ to $L_0/\gamma$. 
The time ${\bf T_B}$ on the solid curve represents the end of acceleration for the back end of the rod, and ${\bf T_F}$ for the front end.\\   
\\

The motion of a rigid rod of moving length $L_0/\gamma$ undergoing constant deceleration from an initial velocity $-V$ to come to rest  at $t=0$ 
is given by the same equations  (\ref{eq:xf}) and (\ref{eq:xb}) as for acceleration, but with the changes $t\rightarrow -t$,
$v\rightarrow -v$, and the interchange of the subscripts $F$ and $B$.  
This corresponds to the reverse motion with time going from $-t$ to 0.  This can be 
depicted on Fig.\ 1, by just moving down the vertical time axis (now thought of as $-t$).
The rod moves with velocity $-V$ until time $-t_B$ (located at $T_F$ in the figure), at which time   
the new back end will start to decelerate, while the front end will continue at constant velocity $-V$, until it starts to decelerate at $-t_F$ (located at $T_B$ in the figure).  
Each end will come to rest at $t=0$, with the length of the rod now $L_0$.  

\section{Impulsive acceleration}
\label{sec:impulse}

Impulsive acceleration occurs when one end of a rod is given an infinite acceleration in an infinitesimal  
time $\Delta t$ so that, in the limit $\Delta t\rightarrow 0$, the product 
$a\Delta t$ approaches a finite change $\Delta V$ in the velocity of one end of the rod.  We consider the case of a rigid rod originally at rest for the which the back end acquires a velocity $V$, and continues to move at that constant rate.   It does not matter where on the rod the impulse is exerted.  Because of the cooperative nature of rigid body acceleration, it will always be the back end that acquires the instantaneous velocity $V$ with  $a'_B\rightarrow\infty$.

 We see from Eq.\ (\ref{eq:lfb}) that, with $a'_B\rightarrow\infty$, the front end will have a finite acceleration $a'_F=1/L_0$.  Then, using Eq.\ (\ref{eq:xf}), the front end of the rod will follow the trajectory 
\begin{eqnarray}
x_F&=&L_0+\left(\sqrt{1+ t_F^2/L^2_0}-1\right)L_0\nonumber\\
&=&\sqrt{L_0^2+t_F^2}.
\label{eq:imp}
\end{eqnarray}
This acceleration will continue until the front end reaches the same velocity as the back end.  From Eq.\ (\ref{eq:tlt}), we see that this occurs at a time
\begin{equation}
T_F=V\gamma L_0,
\label{eq:tf}
\end{equation}
after which both ends continue at the constant velocity $V$.
This impulsive motion is shown as the dashed trajectory in Fig.\ 1 for the same final velocity $V=0.6$ as we used for continuous acceleration.

\section{Rigid body collisions}

The inelastic collision of a rigid rod with a brick wall so that the rod comes to rest after impacting the wall corresponds to moving down in time on the 
dashed trajectory in Fig.\ 1.
The front end of the rod continues at constant velocity $-V$ until it strikes the wall.  The back end starts to decelerate at the time shown as $T_F$ in Fig.\ 1, and
 follows the equation $x=\sqrt{t^2+L_0^2}$ with $t^2$ decreasing until it equals zero and the length of the rod is $L_0$.
Viewers in system S may be surprised to see the back end of the rod start to decelerate before the front end hits the wall.
However, in the rest system of the rod, the onset of deceleration occurs at the same time for each end.
Because the invariant separation of the front and back ends is space-like, the relative time order can be different in other Lorentz frames, but this has no physical significance. 
The early deceleration of the back end seen by viewers in system S is illusory.

An elastic collision of a rigid rod with a wall so that the rod rebounds with the same velocity as it approached the wall corresponds to the same approach to instantaneous 
rest as above, followed by immediate impulsive acceleration as in the previous section.  The collision will be elastic if the final velocity has the same magnitude as the approach velocity.  A partially inelastic collision would occur with a final velocity smaller than the approach velocity.

The collision of two rigid rods along a common line, with no ensuing rotation, can be treated using the preceding formalism.
We consider a collision between two rods of masses $M_1$ and $M_2$ and rest lengths $L_1$ and $L_2$, each moving along the x-axis with velocities $V_1$ and $V_2$, respectively.  Rod 2 leads rod 1, and velocity $V_1$ is greater than $V_2$ (which may be zero or negative) so that the two rods eventually collide.  The right end of rod 1 and the left end of rod 2 have an impulsive impact resulting in final velocities $V'_1$and $V'_2$.  These two ends continue at these constant velocities, which eventually become the final velocities of each entire rod.
The final velocities are determined by conservation of momentum, and conservation of energy for an elastic collision, once a final constant velocity is reached for all parts of each rod.

The leading end of each rod (the right end of rod 1 and the left end of rod 2) move at the constant velocities $V_1$ and $V_2$ until they impact at their common origin of coordinates $x=0$, $t=0$
The original back ends of each rod (the left end of rod 1 and the right end of rod 2) move at these constant velocities until times given by 
Eq.\ (\ref{eq:tf}).  That is
 \begin{eqnarray}
T_1&=&-\gamma_1 V_1 L_1\nonumber\\
T_2&=&\gamma_2 V_2 L_2.
\label{eq:t12}
\end{eqnarray}
These ends then follow Eq.\ (\ref{eq:imp}) so
\begin{eqnarray}
x_1&=&-\sqrt{L_1^2+t_1^2}\nonumber\\
x_2&=&\sqrt{L_2^2+t_2^2}
\label{eq:imp12}
\end{eqnarray}
until times given by 
 \begin{eqnarray}
T'_1&=&-\gamma'_1 V'_1 L_1\label{eq:t'1}\\
T'_2&=&\gamma'_2 V'_2 L_2,
\label{eq:t'12}
\end{eqnarray}
For times greater than $T'_1$ and $T'_2$, each end of each rod continues at constant velocities  $V'_1$ and $V'_2$.
Because the combination $x^2-t^2$ is invariant, equation (\ref{eq:imp12}) holds in any Lorentz system, as long as the impact occurs at time
$t=0$ in that system.

\section{An elastic collision}

In this section, we treat in detail the collision of a rod of rest length $L_0$ with mass $M_1$ and velocity $V=0.6$ ($\gamma_V=1.25$)  with a rod of the same rest length, and mass $M_2$, which is originally at rest.   
We consider two examples, case I with $M_2=2M_1$ so the incoming rod rebounds, and case II with $M_1=2M_2$ where both rods continue in the forward direction. 
We find the final velocities by transforming to the barycentric system for the impact, and then transforming back to the original system.

The velocity to transform to the barycentric system is given by
\begin{equation}
u=\frac{p_1}{E_1+M_2}=\frac{M_1V\gamma_V }{M_1\gamma_V +M_2}=0.231(0.429),\quad \gamma_u=1.028(1.107). 
\end{equation}
In this and subsequent equations the numerical result for case I is given first, followed by the result for case II in parentheses.
The velocity of each rod in the barycentric system is 
\begin{eqnarray}
\overline V_1&=&\frac{V-u}{1-uV}=0.429(0.231),\quad \overline\gamma_1=1.107(1.028)\nonumber\\
 \overline V_2&=&-u=-0.231(-0.429),\quad \overline\gamma_2=1.028(1.107).
\end{eqnarray}

For an elastic collision in the barycentric system, the velocities after impact are the negative of the initial velocities:
\begin{eqnarray}
\overline V'_1&=&-\overline V_1=-1.107(-1.028)\nonumber\\
 \overline V'_2&=&-\overline V_2=-1.028(-1.107).
\end{eqnarray}
If the collision were inelastic, the final velocities would be determined by
\begin{eqnarray}
\overline\gamma'_1\overline V'_1&=&-\epsilon\overline\gamma_1\overline V_1\nonumber\\
 \overline\gamma'_2\overline V'_2&=&-\epsilon\overline\gamma_2\overline V_2,
\label{eq:epsilon}
\end{eqnarray}
where $\epsilon $ is the relativistic coefficient of restitution.  The appearance of the $\gamma $  factors in Eq.\ (\ref{eq:epsilon}) preserves conservation of momentum.

The next step is to Lorentz transform, with velocity $-u$, the barycentric velocities back to the original system where the second rod was originally at rest.
Initially, the right end of rod 1 and the left end of rod 2 have the constant velocities
\begin{eqnarray}
V_1&=&V=0.6\nonumber\\
V_2&=&0
\label{eq:vs}
\end{eqnarray}
After impact, their velocities will be
\begin{eqnarray}
V'_1&=&\frac{u+\overline V'_1}{1+u\overline V'_1}=-0.220(0.220),\quad \gamma'_1=1.025(1.025)
\nonumber\\
V'_2&=&\frac{u+\overline V'_2}{1+u\overline V'_2}=0.439(0.725),\quad \gamma'_2=1.113(1.452). 
\label{eq:vp}
\end{eqnarray}

The left end of rod 1 will move at constant velocity $V_1=0.6$ until a time $T_1$, which is given by Eq.\ (\ref{eq:t12}) to be
\begin{equation}
T_1=-\gamma_1 V_1 L_0=-.731(-.750)L_0.
\end{equation}
The right end of rod 2 will remain at rest until a time $T_2$ given by
\begin{equation}
T_2=\gamma_2 V_2 L_0=0(0).
\end{equation}
The left end of rod 1 and the right end of rod 2 will then follow accelerated motion
 from the times $T_1$ and $T_2$ until the times 
\begin{eqnarray}
T'_1&=&-\gamma'_1 V'_1 L_0=0.225(-0.226) L_0,\nonumber\\
T'_2&=&\gamma'_2 V'_2 L_0=0.488(1.052) L_0.
\end{eqnarray}

The accelerated motion for each rod follows Eq (\ref{eq:imp12}) between the times $T_1\rightarrow T'_1$ and $T_2\rightarrow T'_2$.
That is
\begin{eqnarray}
x_1&=&-\sqrt{L_0^2+ t^2},\quad -0.731(-0.750) L_0< t<0.225(-0.226) L_0,\nonumber\\
x_2&=&\sqrt{L_0^2+ t^2},\quad 0(0)< t<0.488(1.052) L_0.
\label{eq:declab}
\end{eqnarray}
After time $T'_1$, the left end of rod 1 will move at the same constant velocity, $V'_1=-0.220(+0.220)$, as the right end.
After time $T'_2$, the right end of rod 2 will move at the same constant velocity, $V'_2=0.439(0.725)$, as the left end.

The trajectories of the two rods are shown in Fig.\ 3 for case I: $M_2=2M_1$, and in Fig.\ 4
for case II: $M_1=2M_2$.
\begin{center}
\includegraphics[width=4in]{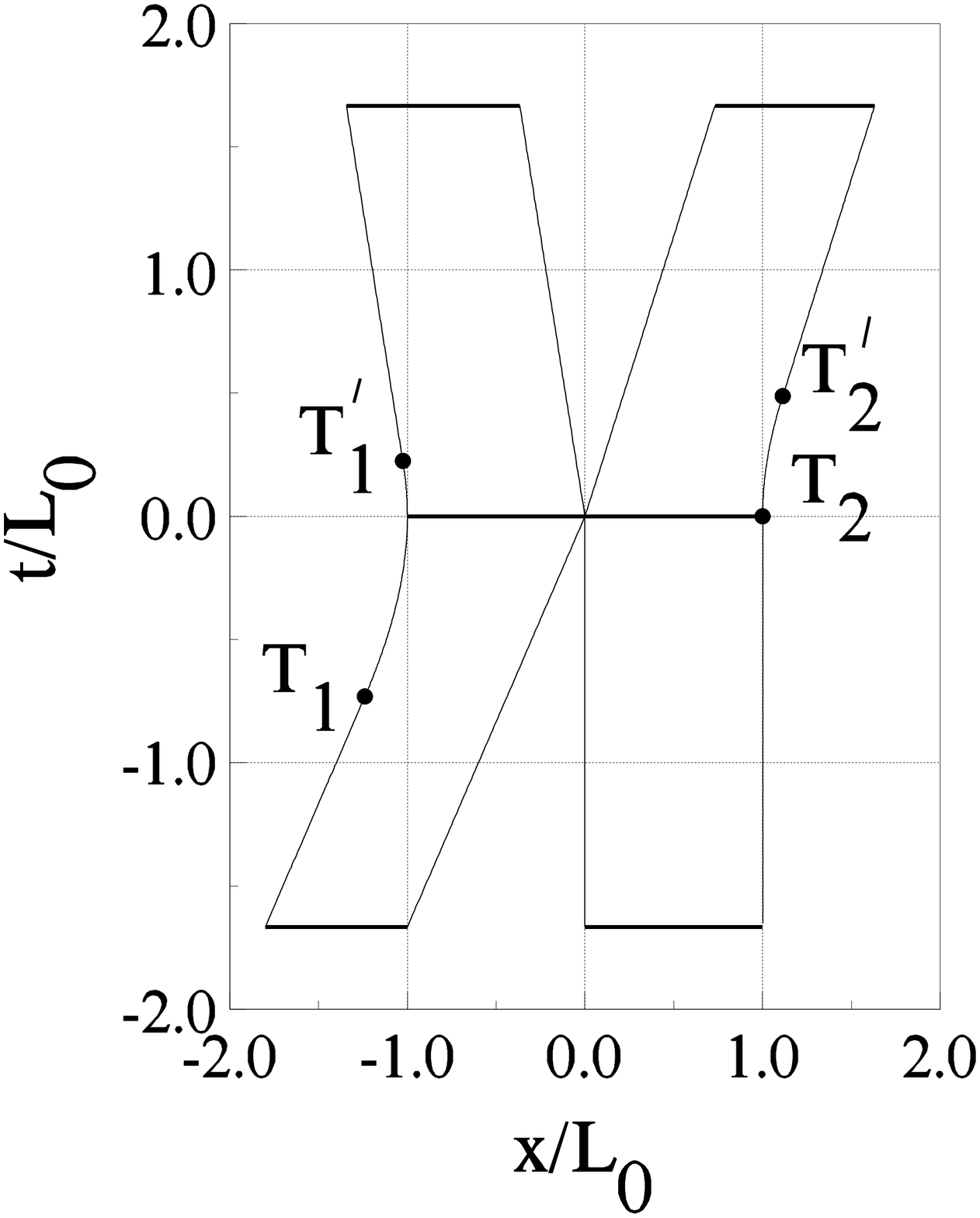}
\end{center}
\noindent
{\bf Fig. 3:}  Trajectories for a collision between rigid rods for case I: $M_2=2M_1$.
The times ${\bf T_1}$ and ${\bf T_2}$ represent the start of acceleration,
and ${\bf T'_1}$ and ${\bf T'_2}$ the end of acceleration for the outer ends of the rods.   \\   
\\
\begin{center}
\includegraphics[width=4in]{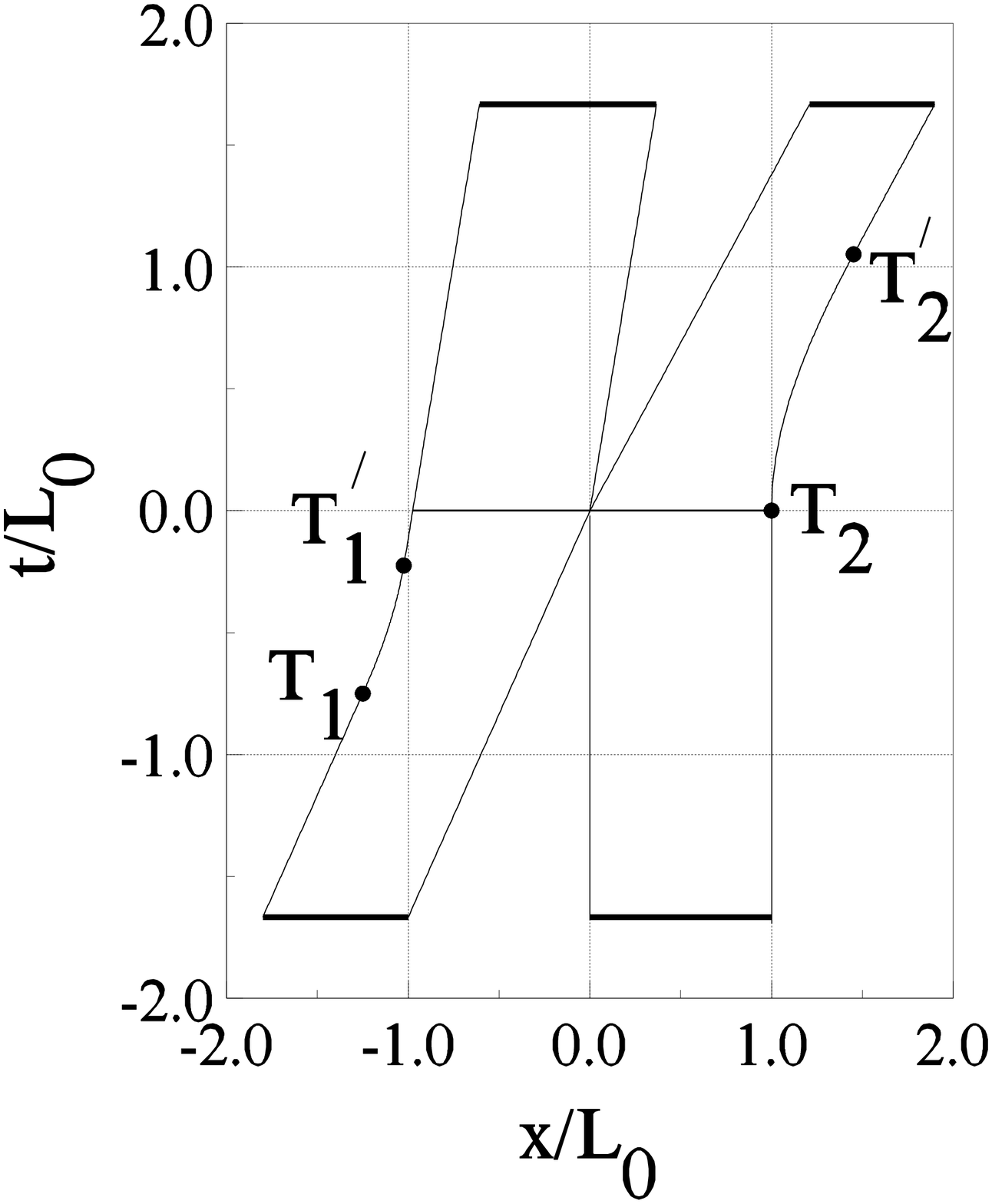}
\end{center}
\noindent
{\bf Fig.\ 4:}  Trajectories for a collision between rigid rods for case II: $M_1=2M_2$.
The times ${\bf T_1}$ and ${\bf T_2}$ represent the start of acceleration,
and ${\bf T'_1}$ and ${\bf T'_2}$ the end of acceleration for the outer ends of the rods.\\   
\\
Rod 2 remains stationary until it is struck by the right end of rod 1.  Then it moves to the right just like the stationary bar given an impulse in 
Sec.\ \ref{sec:impulse}.  The right end of rod 1 and the left end of rod 2 follow the same constant velocity trajectories that point objects of the same mass would follow if energy and momentum conservation were implemented.  The left end of rod 1 follows the curved trajectory shown between times $T_1$ and $T'_1$, and the right end of rod 2 follows its curved trajectory between times $T_2$ and $T_2'$.  After that, they move at the same constant velocities as the other end.

For case I, the lighter rod rebounds and its final trajectory is the same as that of a rod with an impulsive start from rest.
For case II, the heavier rod continues in the forward direction.  
Although the equal time lengths of the two rods are changing as they move, this change is necessary to keep their rest frame lengths constant, as is required for rigid bodies.  
The momentum of either rod is not defined during the accelerated motion, but overall momentum is conserved for the final constant velocities.

\section{Summary}

Using the definition that {\em a rigid body retains its rest frame length  while in motion} we have discussed the accelerated motion and collisions of rigid bodies in special relativity.  We have restricted our treatment to constant accelerations, so as to be able to give simple equations for the trajectories.  
We believe that the general features we have found would also hold for time dependent acceleration, although the curves for the accelerated portions would be somewhat different.  These general features are:
\begin{enumerate}
\item Different parts of an accelerating rigid body undergo different accelerations in the rest frame.
\item In a rigid body collision, the ends that make impulsive contact follow constant velocity paths determined by conservation of momentum and energy
(or appropriate energy loss for an inelastic collision).
\item The outer ends in a collision of a rigid body follow curved, accelerating trajectories in the transition from the initial velocity to the final velocity.
For constant acceleration in the rest frame, the accelerated trajectories are given in the body of this paper.  
\end{enumerate}


\begin{thebibliography}{9}
\bibitem{pauli}W. Pauli, {\it Theory of Relativity} (Pergamon Press, Oxford, UK, 1958) p. 132 .
\bibitem{pp}W. K. H. Panofsky and M. Phillips, {\it Classical Electricity and Magnetism, 2nd Ed.}  (Addison-Wesley, Reading, MA, 1962) p. 287.
\bibitem{ae}A. Einstein, Annalen der Physik, {\bf 17}  891, 1905.
\bibitem{born}M. Born, Annalen der Physik, {\bf 335} 1, 1909.
\bibitem{gh}G. Herglotz, Annalen der Physik, {\bf 336} 393, 1910.
\bibitem{fn}N. Noether, Annalen der Physik, {\bf 336} 919, 1910.
\bibitem{cs}G. Cavalleri and A. G. Spinella, Nuovo Cimento, {\bf B66} 11, 1970.
\bibitem{cbk}C. B. Kafada C, Annalen der Physik, {\bf 484} 325, 1973.
\bibitem{gron}O. Gron, Am. J. Phys., {\bf 45}, 65, 1977.
\bibitem{nic}H. Nikolic, Am. J. Phys. {\bf 67}, 1007, 1999.
\bibitem{pt}F.M. Paiva, A.F.F. Teixeira, arXiv:1201.0670, 2012.
\bibitem{jf}J. Franklin, Eur. J. Phys., {\bf 31} 291, 2010 .
\bibitem{bell}J. S. Bell, {\it Speakable and Unspeakable in Quantum Mechanics} 
(Cambridge University Press, Cambridge, UK, 1st paperback edition, 1973)  p. 67.
\bibitem{jftext}J. Franklin, {\it Classical Electromagnetism}
(San Francisco, CA: Addison-Wesley, 2005).
\end{thebibliography}
\end{document}